\UseRawInputEncoding
\documentclass[twocolumn,aps,prl,longbibliography,10pt]{revtex4-2}
\usepackage{natbib,graphicx,dcolumn,color,lipsum,hyperref}
\usepackage{amsmath,amssymb,verbatim,moreverb,bm,framed}
\usepackage{siunitx}
\usepackage{xcolor}
\usepackage{float}
\definecolor{purple}{rgb}{0.5, 0.0, 0.5} % Purple color
\usepackage[caption=false]{subfig}

\definecolor{orange}{RGB}{255,165,0}

\renewcommand{\phi}{\varphi}
\def\up{\uparrow}
\def\down{\downarrow}
\def\be{\begin{equation}}
\def\ee{\end{equation}}
\def\ber{\begin{eqnarray}}
\def\eer{\end{eqnarray}}

\def\bp{{\bf p}}

\def\bk{{\bf k}}

\def\nn{\nonumber}

\newcommand\commentout[1]{}

\DeclareMathOperator{\sgn}{sgn}
\def\be{\begin{equation}}
\def\ee{\end{equation}}
\def\ber{\begin{eqnarray}}
\def\eer{\end{eqnarray}}
\def\nn{\nonumber}
\def\bk{{\bf k}}

\newcommand{\ie}{{\it i.e.~}} 	%i.e.
\newcommand{\eg}{{\it e.g.~}} 	%e.g.
\newcommand{\tr}{\tilde{r}}
\newcommand{\tz}{\tilde{z}}
\newcommand{\tSigma}{\tilde{\Sigma}}

\begin{document}

\title{
    Hearing the shape of a Dirac drum: Dual quantum Hall states on curved surfaces}
\author{
Ioachim Dusa\(^{1,2}\)\(^{, a}\), 
Denis Kochan\(^{3,4}\)\(^{, b}\), 
Maximilian F\"urst\(^{2}\), 
Cosimo Gorini\(^{5}\)\(^{, c}\), 
Klaus Richter\(^{2}\)\(^{, d}\)}

\affiliation{
\(^1\)TCM Group, Cavendish Laboratory, Department of Physics, Cambridge CB3 0HE, United Kingdom \\
\(^2\)Institut f\"ur Theoretische Physik, Universit\"at Regensburg, 93040 Regensburg, Germany \\
\(^3\)Department of Physics and Center for Quantum Frontiers of Research and Technology (QFort), National Cheng Kung University, Tainan 70101, Taiwan \\
\(^4\)Institute of Physics, Slovak Academy of Sciences, 84511 Bratislava, Slovakia\\
\(^5\)SPEC, CEA, CNRS, Université Paris-Saclay, 91191 Gif-sur-Yvette, France
}

\email{
\text{$^a$}igd27@cam.ac.uk,
\text{$^b$}denis.kochan@phys.ncku.edu.tw,
\text{$^c$}cosimo.gorini@cea.fr,
\text{$^d$}klaus.richter@ur.de
}

\date{\today}

\begin{abstract}
The geometry of a physical system is intimately related to its spectral properties, a concept colloquially referred to as ``hearing the shape of a drum''.
Three-dimensional topological insulator nanowires in a strong magnetic field $B$ generally host Dirac-type quantum Hall (QH) surface states.  The surface itself is shaped by spatial variations of the wires' cross section, yielding a curved geometrical background, the ``drum'', with imprints in the corresponding QH spectra.  We show that the latter are composed of two different classes.  The first one is asymptotically insensitive to the surface shape, scaling as $B^{1/2}$, like regular planar QH states.  Instead, the second has an asymptotic $B$-field dependence intimately related to the wire geometry.
We further demonstrate that an (axial-symmetric) curved nanowire surface possesses a reciprocal partner surface, such that the respective QH spectra are dual to each other upon exchanging angular momentum and magnetic flux.  
Notably, a cone-shaped nanowire, and the Corbino geometry as its limiting case, has a reciprocal partner with a dual QH spectrum that is  $B$-field {\it independent}, with corresponding non-magnetic QH-type states.   
We support our analytical findings by numerical results for $B$-field ranges and wire geometries within reach of current experiment.

\end{abstract}

\maketitle

The idea that the geometry of a physical system is intimately related to its spectral properties was popularised almost 60 years ago, when it was asked whether one ``can hear the shape of a drum'' \cite{kac1966}.  
This essentially refers to the (frequency) spectra of confined waves, whether classical sound waves on a drumhead, light waves in dispersive media or electronic waves in a solid.  In the latter case basic examples are the different electronic spectra of non-interacting particles in flat space, confined within a billiard-type  potential well, in a harmonic trap, in some complex gratings or atomic corral potentials.  A more spectacular one is the formation of Landau Levels (LLs), when freely dispersing electrons subject to a strong magnetic field coalesce into a discrete set of field-dependent quantum Hall states -- \eg of Schr\"odinger or Dirac type \cite{yoshiokabook,mclure1956}.  In these examples the background (the ``shape'') is set by keeping its flat metric fixed but varying the potential.  
The quantum Hall phase is however an ideal platform to consider the more subtle role of metric deformations. This is notably because the latter formally allow to probe the viscosity of a quantum Hall droplet \cite{avron1995} and so-called gravitational anomalies~\cite{can2016}, 
\ie a breakdown of quantum symmetries caused by variations of the space(-time) geometry \cite{avron1995, read2011, abanov2014, klevtsov2015, can2016, can2017}.
This requires  determining the LL spectrum on a curved surface, \eg that of a sphere \cite{haldane1983,Schliemann2008,lee2009,greiter2018} or of a pseudosphere \cite{COMTET1987,GROSCHE1990, Le_2019}.  Colloquially, how a quantum Hall droplet adapts to a curved surface.

\begin{figure}[htbp]
\includegraphics[width=\columnwidth]{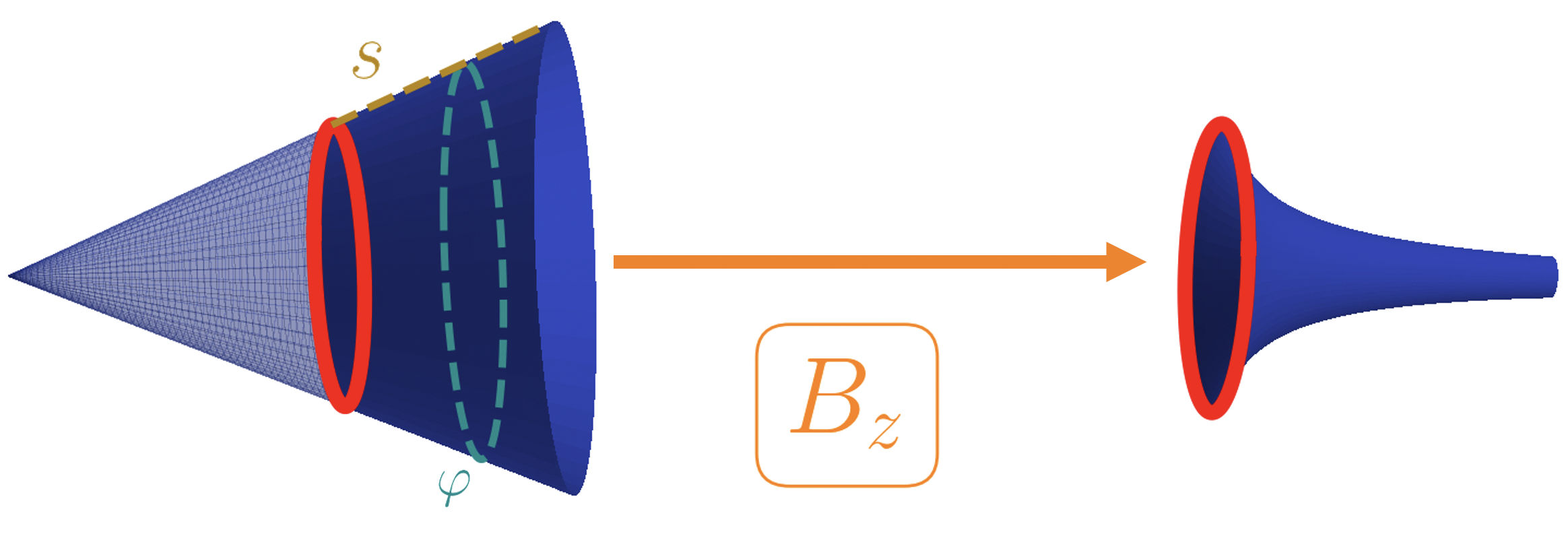}
    \caption{{\bf Reciprocal surfaces} -- Nanocone with opening angle $\theta$ (left) and corresponding reciprocal surface (right) in presence of a coaxial magnetic field of strength $B \!=\! {B_z}$. 
  The leftmost part of the nanocone (transparent) is excluded to avoid the singularity.  The remaining regular surface is defined setting a minimal physical radius, \(\rho_{\text{min}} \equiv a\) (defining the red circle), which simultaneously corresponds to the maximal radius of the reciprocal surface, \(\tilde{\rho}_{\text{max}} = a\). As shown on the left, an arbitrary general surface of revolution $\Sigma$ is parameterised in terms of arc length $s$ and azimuthal angle $\varphi$ (dashed lines), representing the 2D surface hosting metallic Dirac states wrapped around the insulating 3D bulk modelled as an effective vacuum. The axis of rotation is chosen to be parallel to $z$.
    }
    \label{fig:reciprocal_transformation-reworked}
\end{figure}

We recently considered the case of the integer quantum Hall phase for Dirac electrons on surfaces of constant negative curvature \cite{fuerst2024}.  The latter belong to a more general class of surfaces which naturally emerges in shaped 3-dimensional topological insulator (3DTI) nanowires \cite{kozlovsky2020, graf2020}, \ie nanowires made of materials whose bulk is insulating while the surface hosts Dirac-like metallic states \cite{hasan2010colloquium}.  When such systems are immersed in a homogeneous magnetic field in the Tesla range, quantum Hall states form on the metallic surface, the bulk remaining electronically hollow, see Fig.~\ref{fig:reciprocal_transformation-reworked}.  Current experimental capabilities suggest that the realization of such nanowires is challenging but within reach, since 3DTIs can be smoothly shaped at the nanoscale \cite{kessel2017,ziegler2018,behner2023}.  
%%%%%%%%%%%%%%%%%%%%%%%%%%%%%%%%%%%%%%%%%%%%%%%%%%
%%%%%%%%%%%%%%%%%%%%%%%%%%%%%%

Can one thus tell the curved shape of a 3DTI nanowire just by looking at its LL spectrum?  Pursuing the analogy with Kac's drum \cite{kac1966}, we ask whether deforming the drumhead, the shape of the nanowire, set by a metric,  affects the "soundwaves", the Dirac quantum Hall states, in a unique way.  The answer turns out to be multi-faceted.

To be definite we consider axially-symmetric nanowires.  Their outer surface $\Sigma(s,\varphi)$ is parametrized by the arc length $s$ and the azimuthal angle $\varphi$ as shown in the left panel of Fig.~\ref{fig:reciprocal_transformation-reworked}: The radius $r(s)$ and length $z(s)$ are smooth functions satisfying $[r^\prime(s)]^2 + [z^\prime(s)]^2 = 1$, such that points on $\Sigma$ can be cast as $x=r(s)\cos\varphi$, $y=r(s)\sin\varphi$ and $z=z(s)$.  For mathematical convenience we work with rescaled dimensionless quantities $s\equiv l/a$ and radius $r(s)\equiv \rho(s)/a$, with $l, \rho(s)$ the physical arc length and radius and $a \equiv \underset{s}{\min} \{\rho(s)\} > 0$ the minimal radius of the nanowire.\footnote{Other choices for $a$ are possible and make no conceptual difference.}  We assume non-singular surfaces, as exemplified in Fig. \ref{fig:reciprocal_transformation-reworked}, and find two qualitatively different spectral branches, 
{$E^f_\Sigma, E^g_\Sigma$}, for each surface.
%$E_{1/2}, E_r$.  
The first one has universal properties: it is asymptotically blind to the ``drumhead'' shape and scales with $\sqrt{B}$ as relativistic LLs in flat space -- hence the superscript ``$f$'' -- independently of the nanowire surface.  On the contrary the second branch is geometry-sensitive, as highlighted by the ``$g$'' superscript:
\begin{equation}
\label{eq_branches_intro}
    E_\Sigma^f \underset{B \to \infty}{\propto} B^{1/2},
    \quad
    E_\Sigma^g \underset{B \to \infty}{\propto} B^{\eta_{\Sigma}},
\end{equation}
with $\eta_{\Sigma}$ an exponent that depends on the geometry of the nanowire $\Sigma$.  

The geometry-sensitive branch serves as a potential hallmark of the nanowire's shape. Moreover, notably, this branch can be mapped onto a branch of another surface, referred to as the "reciprocal" surface: In particular, for a nanowire with surface \( \Sigma \), there exists a corresponding reciprocal surface \( \tilde{\Sigma} \), characterized by a geometry exponent satisfying \( \eta_{\tilde{\Sigma}} = \frac{1}{2} - \eta_{\Sigma} \).

The behavior of reciprocal surfaces is related to an underlying spectral duality of the 2D Dirac Hamiltonian, where eigenstates of one surface can be systematically associated with those of its reciprocal counterpart. Understanding this duality is crucial for interpreting the relationship between geometry and quantum states.
Note that only a restricted form of duality applies in the Schr\"odinger case, valid only for a specialized class of surfaces (see Supplementary Material \cite{suppmat}).

\paragraph{Dirac Hamiltonian on curved 2D surfaces}
\hspace{-0.25cm}\textemdash 
We write the effective Hamiltonian for the nanowire surface $\Sigma$ \cite{Xypakis_2020, graf2020} in rescaled variables
\begin{equation}
\label{eq_rescaled_H}
    \hat{H} = \frac{\hbar v_F }{a}\bigg [\sigma_1 \bigg(\hat{k}_s - \frac{i}{2}\frac{r'(s)}{r(s)}\bigg) + \sigma_2 \Big(\hat{k}_\phi + \frac{\Phi}{\Phi_0} r(s)\Big)\bigg] ,
\end{equation}
where $\hat{k}_s = -i \partial_s$,  $\hat{k}_\phi = - i \partial_\phi / r(s)$, $v_F$ is the Fermi velocity, and $\sigma_{\{1, 2, 3\}}$ are the corresponding Pauli matrices. 
The coaxial magnetic field along $z$ enters via minimal coupling, $k_\phi \to k_\phi + eA_\phi(s), {\bf A} = A_\phi(s) {\bf e}_\phi =  [B_z a r(s) / 2] {\bf e}_\phi$, written in terms of the magnetic flux $\Phi \!\equiv\! \pi a^2 B_z$ and the flux quantum $\Phi_0\! \equiv\! h / e$, $e>0$. 
We work in the spinor-gauge $|\pm\rangle_{(s,\varphi)} = -|\pm\rangle_{(s,\varphi+2\pi)}$ defined by the outer normal to $\Sigma$ at point $(s,\varphi)$ \cite{fuerst2024}.  Hence $\hat{H}$ acts on $ (\psi_\uparrow,\psi_\downarrow)^\top$, with the spinor field $\psi=\psi_\uparrow(s,\varphi)|+\rangle_{(s,\varphi)} + \psi_\downarrow(s,\varphi)|-\rangle_{(s,\varphi)}$.  Note that $\hat{H}$ is self-adjoint with respect to $\langle \psi|\chi\rangle=\int_\Sigma \mathrm{d}\varphi\mathrm{d}s\,r(s)[\psi^*_\uparrow\chi^{\phantom{*}}_\uparrow + \psi^*_\downarrow\chi^{\phantom{*}}_\downarrow]$.
Since the problem is axially symmetric, the solutions of the Dirac equation are separable, $\psi_{\uparrow/\downarrow}(s, \phi) = e^{{i(m + {1}/{2})\phi} }\psi_{\uparrow/\downarrow}(s)$, with $m \in \mathbb{Z}$ the angular momentum quantum number.  With this ansatz 
  the Dirac equation for Eq.~\eqref{eq_rescaled_H} reads
\begin{align}
  \label{eq_Dirac_super}
    \begin{pmatrix}
			0 & \hat{L}^- \\
			\hat{L}^+ & 0
	\end{pmatrix}
    \psi_\Sigma(s)
    =
    E_\Sigma\psi_\Sigma(s)\quad , \\
    \hat{L}^\pm = i\bigg[ -\partial_s - \frac{1}{2}\frac{r'(s)}{r(s)}  \pm V_\Sigma(s)\bigg],
\end{align}
with energy $E_\Sigma$, arc length wave function $\psi_\Sigma(s) \equiv (\psi_\down(s), \psi_\up(s))^\top$ and effective potential~\cite{kozlovsky2020,graf2020,fuerst2024}
\begin{equation}
  \label{eq_potential}
  V_\Sigma(s) = \frac{m+1/2}{r(s)} + \frac{\Phi}{\Phi_0} r(s) \equiv \frac{\alpha_m}{r(s)} + \beta r(s)
\end{equation}
expressed in units of $\hbar v_F / a$.  
To get a sense of how the spectrum of the Dirac equation behaves in curved geometries,  two analytically solvable cases serve as useful references, that of a cone \cite{kozlovsky2020, suppmat} and of a pseudosphere \cite{fuerst2024}.  In both cases two spectral branches are found \cite{suppmat},
\ber
\label{eq_spectrum_f1}
 \sgn\alpha_m &=& \sgn\beta \Rightarrow E_\Sigma^f(\alpha_m,\beta) \underset{B \to \infty}{\propto} \pm B^{\frac{1}{2}} \, ,
 \\
 \label{eq_spectrum_g1}
 \sgn\alpha_m &\neq& \sgn\beta \Rightarrow E_\Sigma^g(\alpha_m,\beta) \underset{B \to \infty}{\propto} \pm B^{\frac{2-N}{4}},
\eer
with $N = 0, 1$, respectively, for cone and pseudosphere.  These are two first concrete examples of Eq.~\eqref{eq_branches_intro}. 

Before generalizing our result to a wider class of surfaces, let us exploit a remarkable symmetry of the potential $V_\Sigma(s)$
which leads to several important consequences, including the spectral duality of the 2D Dirac Hamiltonian in Eq. \eqref{eq_rescaled_H}.
%%%%%%%%%%%%%%%%%%%%%%%%
\paragraph{Reciprocal surfaces}\hspace{-0.25cm}\textemdash 
Consider the transformation from $\Sigma$, defined by $r(s), z(s)$, to its {\it reciprocal surface} $\tSigma$, given by $\tr(s), \tz(s)$ such that
\begin{equation}
\label{eq_dual_trafo}
\tr(s) \equiv {1}/{r(s)}, \quad \left[\tz'(s)\right]^2 \equiv 1 - \left[ {r'(s)}/{r^2(s)}\right]^2,
\end{equation}
the condition on $\tz$ ensuring that $s$ serves as arc length also for $\tSigma$.
This is illustrated in Fig.~\ref{fig:reciprocal_transformation-reworked} for the surface $\Sigma$ of a cone.
%%%%%%%%%%%%%%%%%%%%%%%%%%%%
The effective potential $V_\Sigma(s)$ is form invariant under Eq.~(\ref{eq_dual_trafo}), \ie
\be
\label{eq_reciprocity}
V_\Sigma(s) = \frac{\alpha_m}{r(s)} + \beta r(s) \to V_{\tSigma}(s) = \alpha_m \tr(s)+ \frac{{\beta}}{\tr(s)} \, ,
\ee
together with the wave function transformation 
\begin{equation}
\label{eq:dual_trafo_wavefunction}
    \psi_\Sigma(s) \to \tilde{r}(s)\psi_{\tilde{\Sigma}}(s),
\end{equation}
preserves the mathematical structure of Eq.~(\ref{eq_Dirac_super}) upon exchanging angular momentum and magnetic flux $\alpha_m \leftrightarrow \beta$.  This notable symmetry does not hold for the Schr\"odinger equation for an arbitrary surface.~\footnote{In curved space, described by a metric tensor $g_{\mu \nu}$ in local coordinates $\{x^\mu\}$, it takes the Laplace-Beltrami form $\Delta =(1/\sqrt {\det g})\partial_{x^i}\left({\sqrt {\det g}}g^{ij}\partial_{x^j}\right)$, which is not form invariant under Eq. ~\eqref{eq_dual_trafo} for an arbitrary $\Sigma$, see Supp. Mat.}
 
Suppose to have the following separable solutions of Eq.~\eqref{eq_rescaled_H} on $\Sigma$ and $\tSigma$ for given $(\alpha_m, \beta)$:
	 \begin{eqnarray}
	  \psi^{\{\alpha_m; \beta\}}_{\Sigma}(s, \phi) &=& e^{i\alpha_m\phi}\psi_{\Sigma}^{\{\alpha_m; \beta\}}(s)
      \;,\;
      E_\Sigma(\alpha_m, \beta),
   \nn\\
    \label{eq:psi_recip}
	\psi^{\{\alpha_m; \beta\}}_{\tSigma}(s, \phi) &=& e^{i\alpha_{{m}}\phi}\psi_{\tilde{\Sigma}}^{\{\alpha_m; \beta\}}(s) \;,\;
    E_{\tSigma}(\alpha_{m}, \beta).
	 \end{eqnarray}
From Eqs.~\eqref{eq_reciprocity}, \eqref{eq:dual_trafo_wavefunction} we conclude  the duality relation
\begin{align}
 \notag
 \psi_{\tilde{\Sigma}}^{\{\alpha_m; \beta\}}(s) &= r(s) \psi_\Sigma^{\{{\beta}; \alpha_{{m}}\}}(s) \\
 E_{\Sigma}(\alpha_m, \beta) &= E_{\tSigma}(\beta, \alpha_m)
 \label{eq_ident_psi_en}
\end{align}
for all $(\alpha_m, \beta)$,
showing that the flux $\beta$ plays the role of the angular momentum $m$ on the reciprocal surface.  Thus the probability densities on the two surfaces can be identified only if $\beta$ is properly chosen:
\begin{equation}
|\psi_{\tSigma}^{\{\alpha_m; \beta\}}(s, \phi)|^2 \!=\! |r(s)\psi_{\Sigma}^{\{\beta; \alpha_m\}}(s, \phi)|^2 ,
\end{equation}
where $\beta \in \mathbb{Z} + \frac{1}{2}$.
We emphasize however that the duality relations \eqref{eq_ident_psi_en} are valid for any value of $\beta$.

%%%%%%%%%%%%%%%%%%%%%%%%%%%%%%

\paragraph{Case study: the nanocone and its reciprocal}\hspace{-0.25cm}\textemdash 
A nanocone is defined by $r(s) = s \sin(\theta / 2)$, see Fig.~\ref{fig:reciprocal_transformation-reworked}, left panel. 
Its spectral branches read \cite{suppmat}
\begin{gather}
 \label{eq:energies_cone_2}
 E_\Sigma^f(\alpha_m, \beta, n) \! = \! \\
 \pm \sqrt{4\Big [\sin\Big(\frac{\theta}{2}\Big)\Big(n\!+\! \frac{1}{2}\Big) \!+\!|\alpha_m| \Big ]|\beta|} \underset{B \to \infty}{\propto} \!B^{\frac{1}{2}} \notag
\end{gather}
for $\sgn\beta = \sgn(\alpha_m)$, and
\begin{equation}
 \label{eq:energies_cone_1}
 E_\Sigma^g(\alpha_m, \beta, n) = \pm \sqrt{4 \sin(\theta / 2)\,|\beta|\, n} \propto B^{\frac{1}{2}} ,
\end{equation}
for $\sgn\beta \neq \sgn(\alpha_m)$, with $n\geq 0$. The latter are Landau levels (highly) degenerate in angular momentum $m$.  Both branches scale as $B^{1/2}$.

%%%%%%%%%%%%%%%%%%%%%%%%%%%%%%%%%%%%%%%%%%%%%%%%5
\begin{figure}[htbp]
    \centering
    \includegraphics[width=\linewidth]{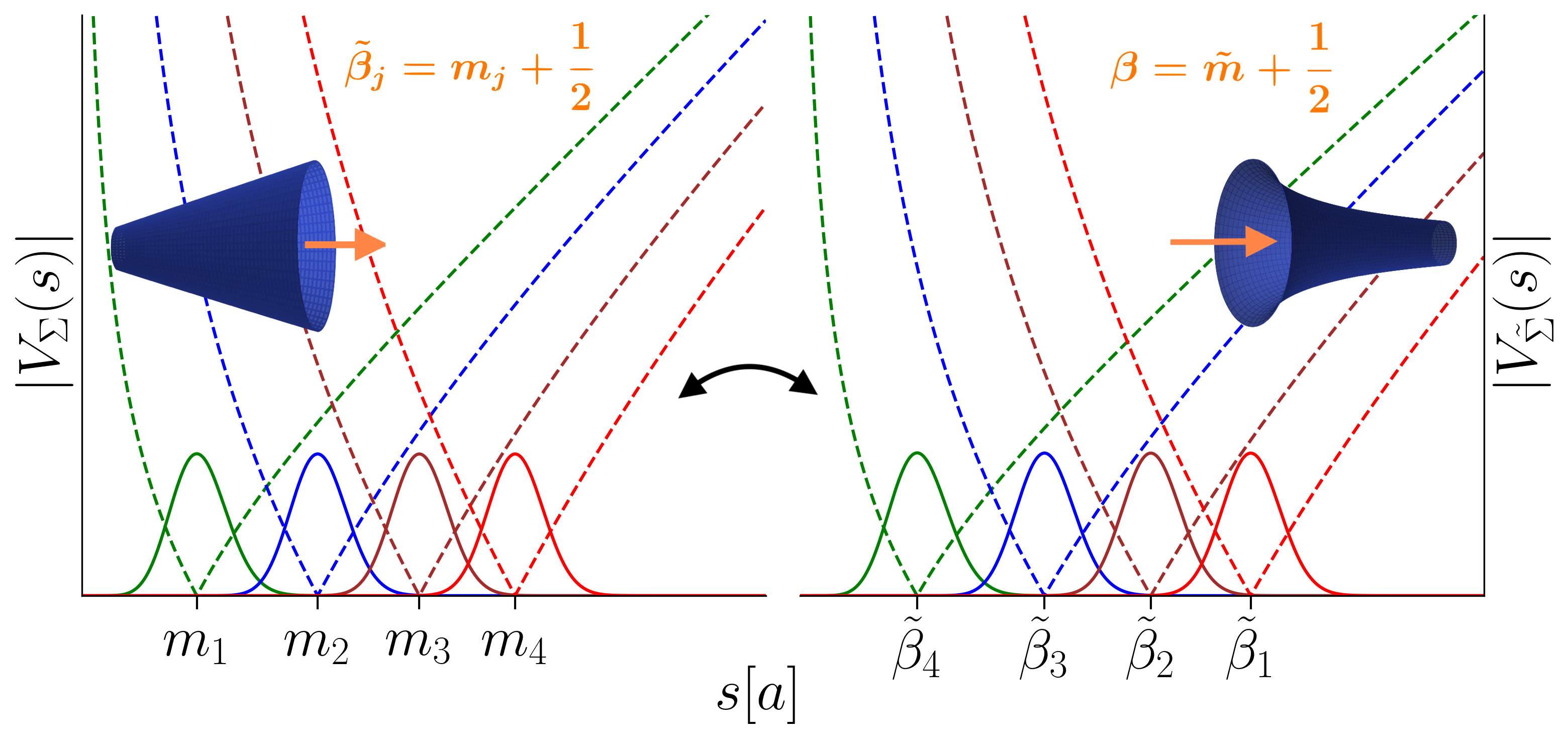}
        \footnotesize
         \caption{{\bf Wave function duality} -- (Ground state) wave functions in the  corresponding potential wedges (Eq.~\ref{eq_reciprocity})) as a function of arc length $s$ for the nanocone (left), for different angular momentum quantum numbers $m_j, j=1\dots 4$, at fixed magnetic flux $\beta \in \mathbb{Z} + 1/2$ to enable direct mapping to the reciprocal surface (right).  There dual wave functions and wedges are labeled by different magnetic fluxes $\tilde{\beta}_j = m_j + 1 / 2, j=1,\dots4$ at fixed angular momentum quantum number $\tilde{m} = \beta - 1/2$.  Dual pairs on the cone are marked with same color code.  Values $m_j = [10, 40, 80, 130]$ and $ \beta = -6.5$ were taken.} 
     \label{fig_double_potential}
%double_potential.py
\end{figure}
%%%%%%%%%%%%%%%%%%%%%%%%%%%%%%%%%%%%%%%%%%%%%%%%%%

Exchanging $\beta \leftrightarrow \alpha_m$, see Eq.~\eqref{eq_ident_psi_en}, one obtains the dual eigenfunctions and eigenenergies on the reciprocal surface. 
 Figure \ref{fig_double_potential} illustrates the dual eigenfunctions, located in the corresponding $s$-dependent effective potential wells. The energies read explicitly
\begin{gather}
\label{eq:recip_cone_energy_2}
	E_{\tSigma}^f(\alpha_m, \beta, n) = \\ 
     \pm \sqrt{4\Big [\sin\Big(\frac{\theta}{2}\Big)\Big(n\!+\! \frac{1}{2}\Big) \!+\!|\beta| \Big ]|\alpha_m|} \underset{B \to \infty}{\propto} \!B^{\frac{1}{2}} \, . \notag
\end{gather}
\begin{equation}
 \label{eq:recip_cone_energy_1}
 E_{\tSigma}^g(\alpha_m, n) \!=\! \pm \sqrt{4\sin(\theta / 2)\, |\alpha_m |n} \propto B^0 \!=\!\text{const.} 
\end{equation}
Most notably, the fact that the Landau fan, Eq. \eqref{eq:energies_cone_1}, is independent of $\alpha_m$, implies that the  dual spectrum, Eq.~\eqref{eq:recip_cone_energy_1}, is $\beta$-independent, \ie $m$-degeneracy 
implies magnetic-field independence at finite $B$-fields! 

%%%%%%%%%%%%%%%%%%%%%%%%%%

\begin{figure}[htbp]
    \centering
\includegraphics[width=\linewidth]{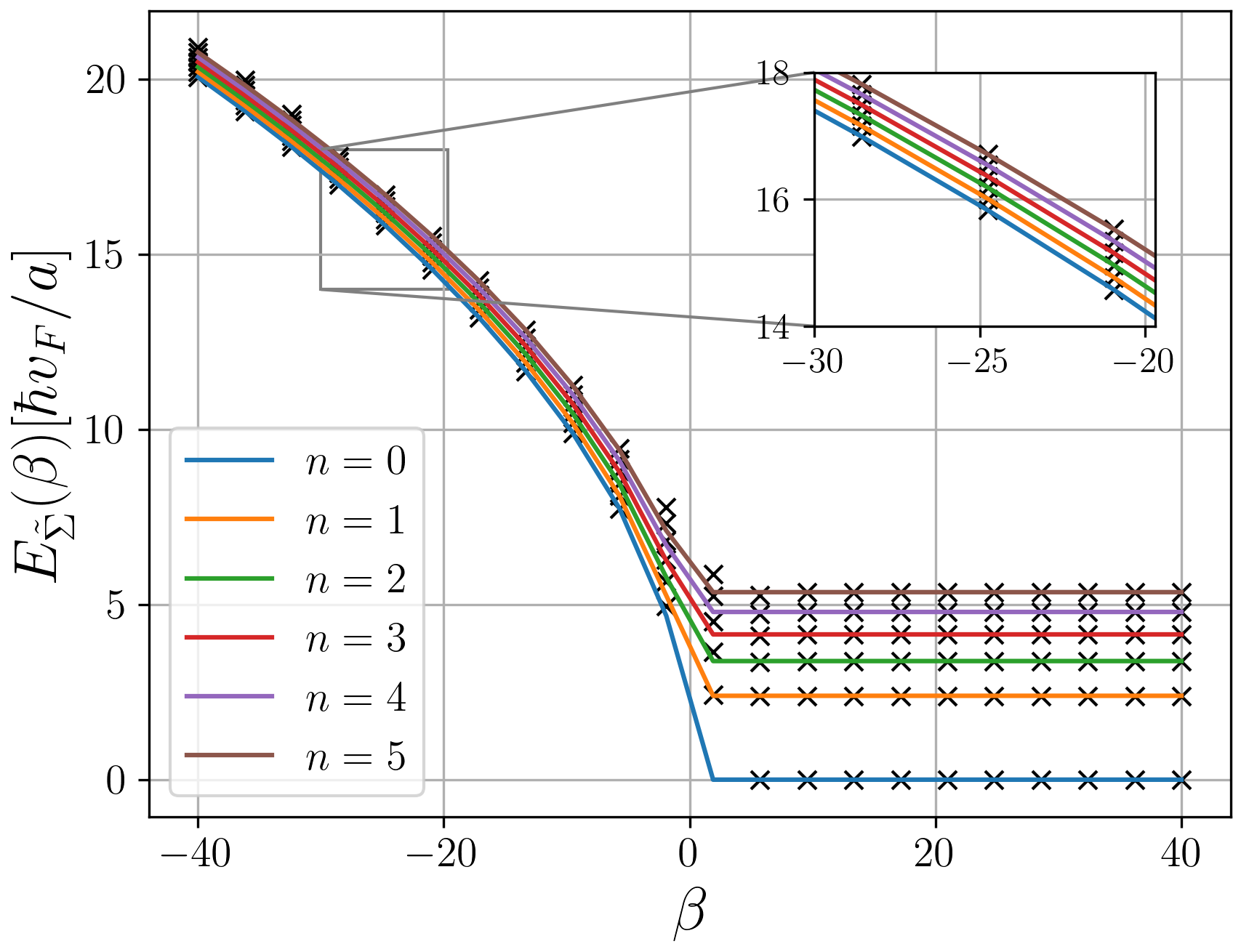}
        \footnotesize
        \caption{{\bf Flat Landau levels of the reciprocal cone} -- Reciprocal cone energy levels, shown as a function of flux
    $\beta=\Phi/\Phi_0$ for angular quantum number $m = -3$ ({\em i.e.} $\alpha_m=-5/2$) and principal quantum number $n=0$ to $5$ (from bottom to top).  The opening angle $\theta \!=\!70^\circ$.  Eqs.~\eqref{eq:recip_cone_energy_1} and \eqref{eq:recip_cone_energy_2} (lines) are compared to numerical results (crosses), obtained with a lattice spacing $\Delta_s = 0.0125 [a]$ in the arc length direction. }
    \label{fig:reciprocal_cone_spectrum_B_field}
%junk.py
\end{figure}
%%%%%%%%%%%%%%%%%%%%%%%%

This is shown in Fig.~\ref{fig:reciprocal_cone_spectrum_B_field} for $m=-3$ and various $n$. Lines are plots of Eqs.~\eqref{eq:recip_cone_energy_2} and \eqref{eq:recip_cone_energy_1}, while crosses show results from a numerical solution of the Dirac equation of the reciprocal cone. Regarding the numerics, in all cases studied we add a small Wilson mass term to the corresponding Dirac Hamiltonian and use hard wall boundary conditions.
Small deviations from the analytical results appear for small $|\beta|$ where the quantum Hall states hit the boundaries.

For $\theta\!=\!\pi$ the nanocone corresponds to the famous Corbino disk geometry \cite{yoshiokabook} (with rescaled inner disk radius $a$) with standard quantum Hall energy levels $ \pm 2(\hbar v_F/a)\sqrt{|\beta|\, n}$.  The (rescaled) dual energies on the reciprocal surface read $\pm 2(\hbar v_F/a)\sqrt{|m+1/2|n} $.

\begin{figure}[htbp]
    \centering
  \includegraphics[width=\linewidth]{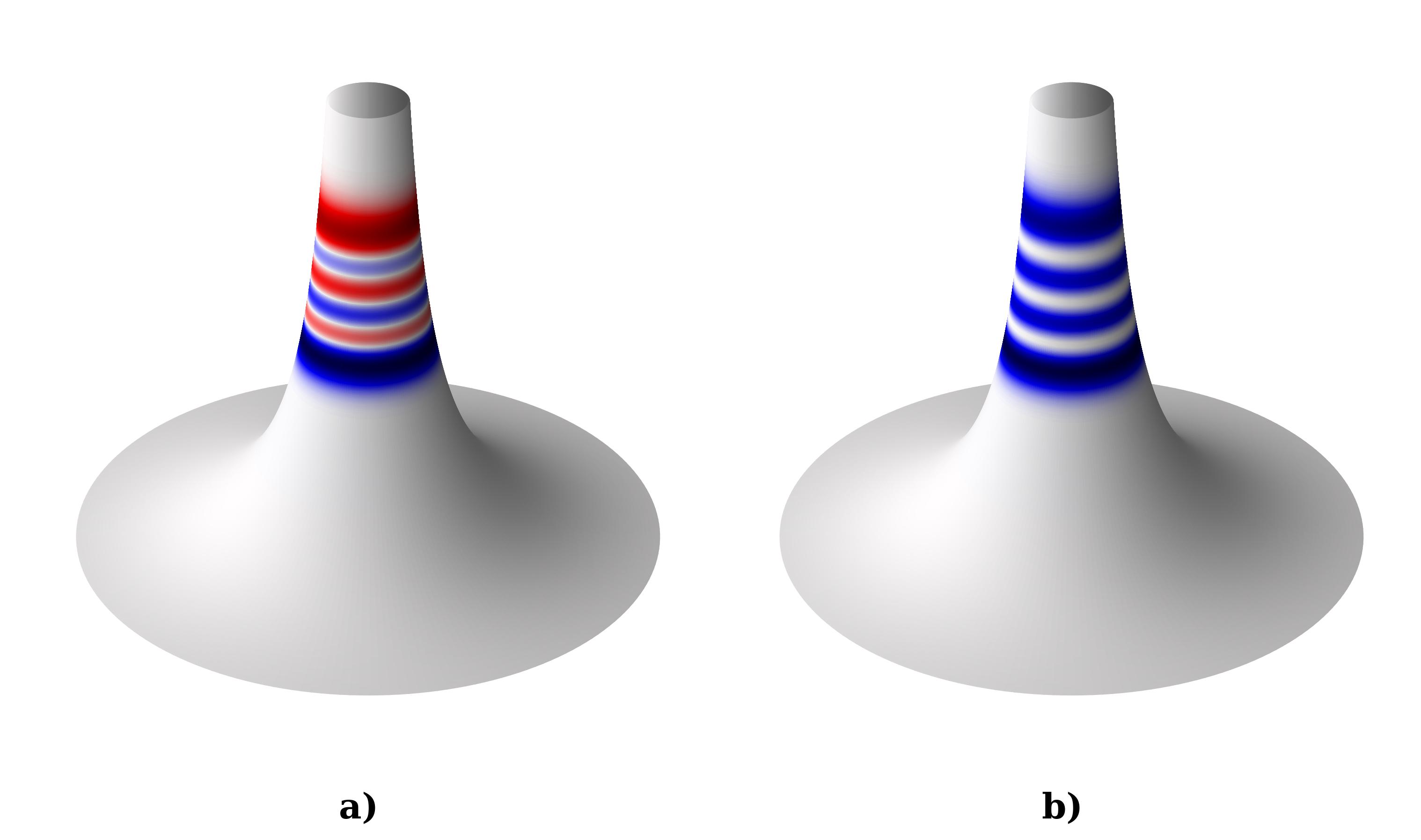}
        \footnotesize
        \caption{{\bf Local angular current densities} $j^{(n)}_\varphi(s)$ on the reciprocal cone for $n=3$ and a) $\sgn(\alpha_m) \neq \sgn(B), m=-3$; b) regular quantum Hall states with $\sgn(\alpha_m) = \sgn(B), m=2$.  Clockwise (counter clockwise) flowing current is shown in blue (red), stronger color corresponding to higher current.
        The magnetic flux is $\beta = 40$, the opening angle of the original cone $\theta = 70^\circ$.}
     \label{fig_density_current_combined}
%bound_states_solver_wilson_new.py
\end{figure}

The fact that the energies are $B$-field independent, see spectrum in
Fig.~\ref{fig:reciprocal_cone_spectrum_B_field} for $\beta > 0$, implies that the corresponding  eigenstates states cannot carry any $z$-magnetic moment $m_z$, since
\begin{equation}
m_z(\alpha_m, n) \propto \partial  E_{\tSigma}^g(\alpha_m,n)/\partial B = 0.
\end{equation}
Indeed the  $E_{\tSigma}^g$-modes do not follow the standard behaviour at the heart of Laughlin's topological argument \cite{laughlin1981}: Rather than flowing into/out of the droplet when a change in $B$ adds/removes flux quanta through the cross-section, they move in space along the axis (see Fig.~\ref{fig:reciprocal_cone_spectrum_B_field}, right) and internally twist to counteract flux changes and thus remain within the droplet keeping $E_{\tSigma}^g(\alpha_m,\beta,n)$ unchanged.  This is shown in Fig.~\ref{fig_density_current_combined}, where the local (angular) current density $j^{(n)}_\phi(s) =  -e v_F \psi_{\tilde{\Sigma}, n}^{\dagger}(s) \sigma_2 \psi_{\tilde{\Sigma}, n}(s)$ is plotted as a function of the arc length coordinate for both spectral branches~\cite{suppmat}. For  $E_{\tSigma}^g$ modes the current changes sign, \ie handedness, while going along the nanowire, resulting in the expected zero overall magnetic moment
\begin{equation}
\label{eq_m_z_n}
    m_z(\alpha_m, n) = \frac{1}{2}\int_\Sigma \mathrm{dV} \bigg( \vec{\rho} \times \vec{j} \bigg)_z = \pi\int \mathrm{d}s \; \rho^2(s) j^{(n)}_\phi(s) = 0.
\end{equation}

%%%%%%%%%%%%%%%%%%%%%%%%%%%%%%%%%%%%%%%%%%%%%%%%%%%%%%%%%%%%%%%%%%%%%%%%%%%%%%%%%%%%%%
%%%%%%%%%%%%%%%%%%%%%%%%%%%%%%%%%%%%%%%%%%%%%%%%%%%%%%%%%%%%%%%%%%%%%%%%%%%%%%%%%%%%

%%%%%%%%%%%%%%%%%%%%%%%%%%%%%%%%%%%%%%%%%%%%%%%%%%%%%%%%%%%%%%%%%%%%%%%%%%%%%%%%%%%%%%%%
\begin{figure}[htbp]
	\centering
\includegraphics[width=\linewidth]{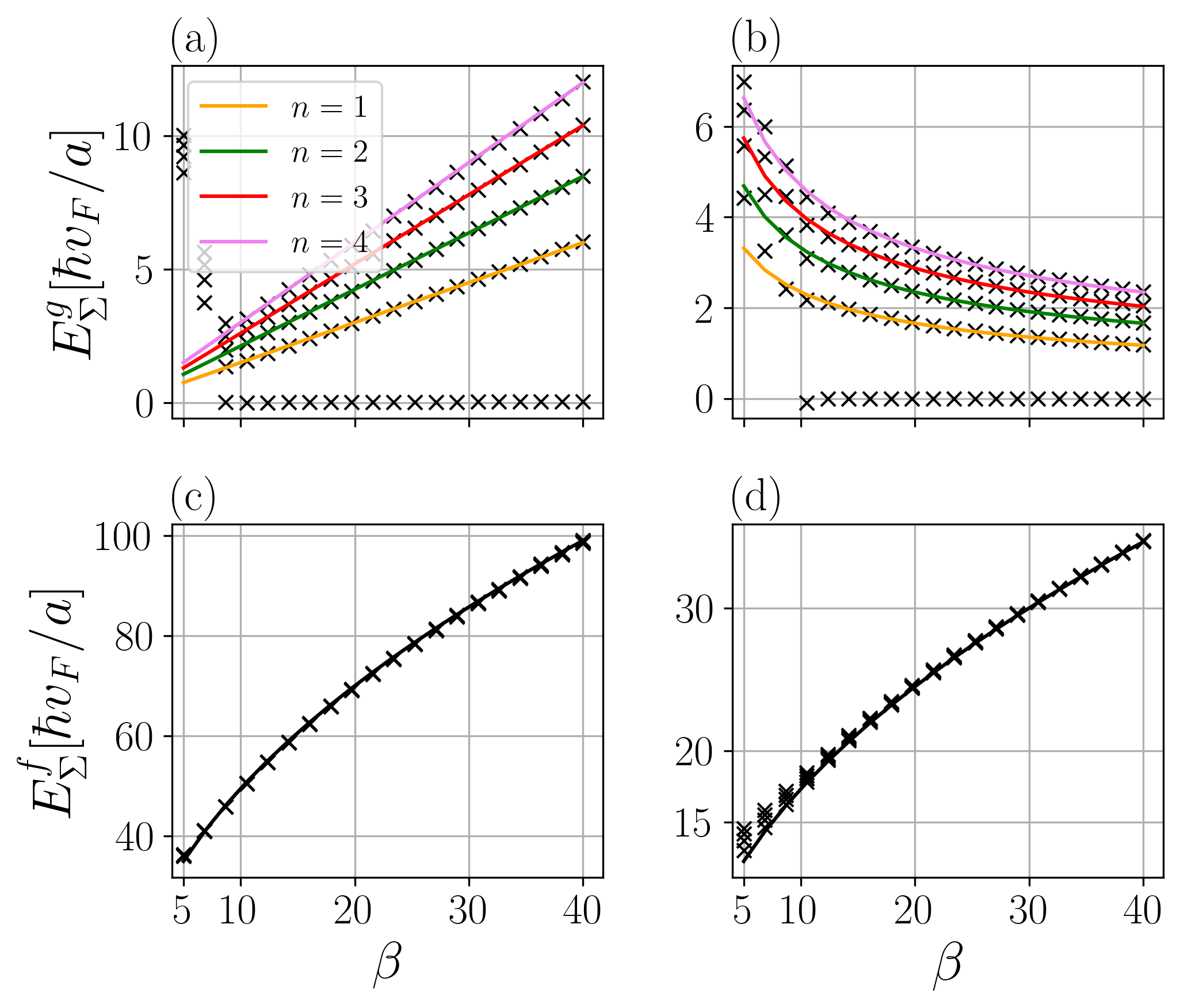}
        \footnotesize
   \commentout{ \caption{
    The plots depict the flux dependence of the energy of different radial profiles with $r' \propto r^{N}$ in the case $\text{sgn}(B) = \text{sgn}(\alpha_m)$ (c, d) and $\text{sgn}(B) \neq \text{sgn}(\alpha_m)$ (a, b) for a fixed angular quantum number $\alpha_m$ in each case. The radial profile equation is trivially solved, and its spectrum is computed numerically, which is shown by the dots. The continuous lines represent the fitting monomial $E(\beta) = q\beta^d$ with the variables having the values: (c) $q = \sqrt{245}$, $d = 1 / 2$, (d) $q = \sqrt{30}$, $d = 1 / 2$, (a) $q = \textcolor{orange}{\sqrt{0.025}}, \textcolor{green}{\sqrt{0.045}}, \textcolor{red}{\sqrt{0.0675}}, \textcolor{purple}{\sqrt{0.090}}$, $d = 1$ (b) $q = \textcolor{blue}{\sqrt{55}}, \textcolor{orange}{\sqrt{110}}, \textcolor{green}{\sqrt{165}}, \textcolor{red}{\sqrt{220}}$, $d = - 1 / 2$
    , confirming our conjecture eq. \eqref{eq_conjecture_1}. We chose for $N = -2$ (a, c) the angular quantum numbers $m=-60,60$ while for $N = 4$ (b, d) we had $m=-7, 7$. We used an arc length lattice spacing of $\Delta_s =0.0125 [a]$ lattice points and calculated for each flux $\beta$ the first $n\leq 4$ energy levels. 
    } }
    \caption{{\bf Hearing the shape of nanowires} -- Flux dependence of the energy levels for radial profiles defined by \( r' \propto r^{N} \), with \( N = -2 \) and (a) \( m = -60 \) and (b) 60 as well as \( N = 4 \) with (c) \( m = -7\) and (d) 7. Hence, \(\text{sgn}(B) \neq \text{sgn}(\alpha_m)\) in the upper panels and \(\text{sgn}(B) = \text{sgn}(\alpha_m)\) in the lower ones. 
The radial equation is solved numerically with arc-length lattice spacing $\Delta_s = 0.0125 [a]$, yielding energy levels for principal quantum numbers $n \leq 4$ as a function of $\beta$, shown as crosses.
Solid lines are fits to the monomial function \( E(\beta) = q\beta^d \), with parameters (clockwise from the upper left panel): \( q = \textcolor{orange}{\sqrt{0.025}}, \textcolor{green}{\sqrt{0.045}}, \textcolor{red}{\sqrt{0.0675}}, \textcolor{purple}{\sqrt{0.090}} \), \( d = 1 \); \( q = \textcolor{orange}{\sqrt{55}}, \textcolor{green}{\sqrt{110}}, \textcolor{red}{\sqrt{165}}, \textcolor{purple}{\sqrt{220}} \), \( d = -\frac{1}{2} \); \( q = \sqrt{30} \), $d = \frac{1}{2}$; \( q = \sqrt{245} \), \( d = \frac{1}{2} \).  Thus, the conjecture Eq.~\eqref{eq_conjecture_1} is numerically confirmed.
}
\label{fig:numerical_evidence}
% B-field-dependence.py
\end{figure}
%%%%%%%%%%%%%%%%%%%%%%%%%%%%%%%%%%%%%%%%%%%%%%%%%%%%%%%%%%%%%%%%%%%%%%%%%%%%%%%%%%%%%%%

\paragraph{A conjecture}\hspace{-0.25cm}\textemdash
The spectral branches
$E_{\tSigma}^f$ and $E_{\tSigma}^g$ on the reciprocal cone, Eqs.~\eqref{eq:recip_cone_energy_2} and  \eqref{eq:recip_cone_energy_1}, scale in accordance with Eqs.~\eqref{eq_spectrum_f1} and \eqref{eq_spectrum_g1}.  Given these insights, we propose a generalization of our conclusions to a broad class of surfaces.  Specifically, we postulate that the splitting of the quantum Hall spectrum into the two branches 
\begin{equation}
\label{eq_conjecture_1}
    E_\Sigma^f \underset{B \to \infty}{\propto} B^{1/2} \quad \text{and} \quad E_\Sigma^g \underset{B \to \infty}{\propto} B^{(2-N)/4}
\end{equation}
holds in fact for arbitrary surfaces of revolution satisfying 
\begin{equation}
 \label{eq_conjecture_0}
 \frac{\mathrm{d}r}{\mathrm{d}s}(s) \propto \left[r(s)\right]^{N},
\end{equation}
for some integer $N \in \mathbb{Z}$.  We numerically validate our conjecture for different geometries, see Fig.~\ref{fig:numerical_evidence} for $N = -2$ and $ 4$, and further cases in \cite{suppmat}.  

Furthermore, for any (regular) surface of revolution the radius derivative  $r'(s)$  can be Taylor-expanded as
\begin{equation}
\label{eq:decomposition}
\frac{\mathrm{d}r}{\mathrm{d}s}(s) = \sum_{N \in \mathbb{Z}} a_N [r(s)]^N ,
\end{equation}
allowing the geometry-dependent asymptotics of the energy $E_\Sigma^g$ to be determined individually for each monomial term $r^N$ using Eq.~\eqref{eq_spectrum_g1}. The dominant contribution from this expansion then governs the spectral scaling for the full surface.

To validate this behavior, we numerically investigated functions beyond the monomial forms discussed in Fig.~\ref{fig:numerical_evidence}, namely transcendental functions, as detailed in  \cite{suppmat}.
For example, for
$r'(s) \propto \sin(r) = r^1 - r^3 / 3! + \dots$, the conjecture correctly predicts the dominant fourth-root behavior ($\eta_\Sigma = 1/4$) for the curvature-dependent spectrum in the strong-field limit.
%%%%%%%%%%%%%%%%%%%%%%%%%%%%%%%%%%%%%%%%%%%%%%%%%%%%%%%%%%%%%%%%%%%%%%%%%%%%%%%%%%%%55

\paragraph{Conclusions}\hspace{-0.25cm}\textemdash
We studied the Dirac QH spectrum of axially symmetric surface states identifying two distinct branches that can be potentially realized in laboratory by surface modes of 3D TI nanowires. First, we showed that one branch -- we call it geometric -- allows to sense the shape of the nanowire while exploring its LL spectrum for large magnetic field. Thus 3D TIs can serve as another platform for ``hearing the shape of drum''. Second, we uncovered a spectral duality of 2D Dirac equation in axial magnetic field -- a sort of transformation of reciprocal radii -- that maps eigenvalues and eigenfunctions between two surfaces whose radii are related by $r(s)\rightarrow \tilde{r}(s)=1/r(s)$. Third, we found a curved surface, the reciprocal cone, with peculiar counter-reaction to $B$.  It carries ``non-magnetic quantum Hall states'', which reorganize upon changes in $B$ in a way to carry zero magnetization and thus canceling the field dependence of the corresponding eigenenergies.
We conjecture these states to exist for generic (smooth) surfaces of revolution beyond the specifics
of 3D TIs, which we verified by numerous numerical calculations.

We conclude by considering some model limitations and material-related aspects.  First, our analysis was limited to orbital effects neglecting the Zeeman coupling.  For coaxial magnetic fields $B_z$ we expect the latter to have the general form
\be
 \label{eq_Zeeman}
 H_Z = \frac{\mu_B}{2}\bigg(\sum_i\,g_i(s,\phi)\sigma_i\bigg) B_z,\quad i=x,y,z, 
\ee
with coordinate-dependent and anisotropic effective $g$-factors. 
As standard in $\bk\cdot\bp$ theory~\cite{winklerbook}, this Zeeman term arises from a folding down procedure and carries information about material-specific anisotropies \cite{liu2010}, including notably the real spin character of surface states \cite{zhang2012,silvestrov2012,brey2014}.  Zeeman couplings in the $x$- and $y$-directions do not spoil the (chiral) structure of Eq.\eqref{eq_Dirac_super}, {\it ergo} our conclusions remain valid.  A $z$-term on the other hand does, so that a reciprocal surface with invariant spectrum cannot be exactly defined as done in Eqs.~\eqref{eq_reciprocity} and \eqref{eq:dual_trafo_wavefunction}. However, the $B$-dependent Zeeman splitting is expected to be small for the field strengths considered.
To what extent the duality is broken and if a different one can still be identified, depends on the geometry and material. 

Finally, we neglected standard non-magnetic disorder, as it does not qualitatively affect our conclusions as long as broadening remains smaller than the resolution of our spectra.  It might play a more interesting role in transport, which however lies beyond the scope of this work.  

\paragraph{Acknowledgements}\hspace{-0.25cm}\textemdash
I.G.D.  thanks the Service de Physique de l'Etat Condensé (SPEC) at CEA Saclay for the warm hospitality and support received during his visit, which contributed to the progress of his Master thesis work at University of Regensburg. C.G. thanks the STherQO members, in particular Arseni Goussev and R\`{e}my Dubertrand, for fruitful discussions.
D.K.~acknowledges partial support from the project 
IM-2021-26 (SUPERSPIN) funded by the Slovak Academy of Sciences via the programme IMPULZ 2021, and thanks Vladim\'{i}r Balek and Mari\'{a}n Fecko for fruitful discussions.
The work was funded by the
Deutsche Forschungsgemeinschaft (DFG, German Research Foundation) within Project-ID 314695032 – SFB 1277 (subproject A07).

\bibliographystyle{apsrev4-2}
\bibliography{references}

%%%%%%%%%%%% SUPP MAT

\clearpage
\onecolumngrid  % Ensure single-column format for clarity

\begin{center}
    \textbf{\large Supplementary Information: \\
    Hearing the shape of a Dirac drum: Dual quantum Hall states on curved surfaces}
\end{center}
\vspace{1cm}

% Reset numbering for Supplementary Material
\renewcommand{\theequation}{S\arabic{equation}}
\renewcommand{\thefigure}{S\arabic{figure}}
\renewcommand{\thetable}{S\arabic{table}}
\renewcommand{\thepage}{\arabic{page}}
\setcounter{equation}{0}
\setcounter{figure}{0}
\setcounter{table}{0}
\setcounter{page}{1}

\twocolumngrid

In the following, we briefly derive a special form of duality in the case of Schr\"odinger electrons, highlighting the restricted nature of Schr\"odinger duality compared to the Dirac case.”

We then derive the analytical bound states on the nanocone, including wave functions and energies. In addition, we present the derivation of the magnetic moment of general surfaces of revolution from the local current density. 

In the last section, we provide further justification for the applicability of the presented conjecture to (almost) arbitrary surfaces of revolution.

\section{\textbf{Reciprocity in the Schr\"odinger case}}
In this chapter, we derive a special duality of Schr\"odinger electrons for a surface of revolution \(\Sigma\) and its reciprocal counterpart \(\tilde{\Sigma}\), which will turn out to only hold for one class of surfaces, namely $r \propto e^{-\frac{c}{2} s^2 - d s}$, where $c, d\in \mathbb{R}$.

Using the Laplace operator in curved space,
\begin{equation}
\Delta = \frac{1}{\sqrt{\det g}} \partial_{x^i} \left( \sqrt{\det g} g^{ij} \partial_{x^j} \right),
\end{equation}
we obtain the Schr\"odinger equation for \(\Sigma\):
\[
\frac{1}{r} \left[ r \psi_\Sigma'(s) \right]' + \left( \frac{\alpha_m}{r} + \beta r \right)^2 \psi_\Sigma(s) = E_\Sigma \psi_\Sigma(s), \quad (\Sigma)
\]
and for its reciprocal surface \(\tilde{\Sigma}\):
\[
\frac{1}{\tilde{r}} \left[ \tilde{r} {\psi_{\tilde{\Sigma}}}'(s) \right]' + \left( \frac{\alpha_m}{\tilde{r}} + \beta \tilde{r} \right)^2 {\psi_{\tilde{\Sigma}}}(s) = E_{\tilde{\Sigma}} {\psi_{\tilde{\Sigma}}}(s), \quad (\tilde{\Sigma})
\]
where we assume a separation ansatz for the wavefunction:
\begin{equation}
 \psi_\Sigma(s, \varphi) = e^{i(m + 1/2)\varphi} \psi_\Sigma(s).
\end{equation}

As in the main text, we impose the transformation relations:
\begin{equation}
 r = \frac{1}{\tilde{r}}, \quad \psi_{{\Sigma}} = \tilde{r} \psi_{\tilde{\Sigma}}, \quad \alpha_m \leftrightarrow \beta.
\end{equation}

Substituting these into the equation for \(\Sigma\) gives:
\begin{equation}
\tilde{r} \left[ \frac{1}{\tilde{r} } (\tilde{r}  \psi_{\tilde{\Sigma}})' \right]' + \left( \beta \tilde{r} + \frac{\alpha_m}{\tilde{r}} \right)^2 \tilde{r} \psi_{\tilde{\Sigma}} = E_\Sigma \tilde{r} \psi_{\tilde{\Sigma}}.
\end{equation}

To ensure the duality between the two equations, the kinetic and potential energy terms must be identical, as the latter remain unchanged under the transformation of the angular quantum number and magnetic flux, \( \alpha_m \leftrightarrow \beta \). This requirement leads to the condition:
\begin{equation}
\frac{1}{\tilde{r}} \left[ \tilde{r} {\psi_{\tilde{\Sigma}}}'(s) \right]' - E_{\tilde{\Sigma}} \psi_{\tilde{\Sigma}}\overset{!}{=} \left[ \frac{1}{\tilde{r} } (\tilde{r}  \psi_{\tilde{\Sigma}})' \right]'  - E_{{\Sigma}} \psi_{\tilde{\Sigma}}.
\end{equation}

Expanding the derivatives results in:
\begin{equation}
 - E_{\tilde{\Sigma}} \psi_{\tilde{\Sigma}} \overset{!}{=} \bigg(\frac{\tilde{r}'}{\tilde{r}}\bigg)' \psi_{\tilde{\Sigma}}- E_{{\Sigma}} \psi_{\tilde{\Sigma}}.
\end{equation}

For this equality to hold, the function \(\tilde{r}\) must satisfy the constraint:
\begin{equation}
\bigg(\frac{\tilde{r}'}{\tilde{r}}\bigg)' = \text{const} \equiv c,
\end{equation}
which also implies a shift in energy:
\begin{equation}
E_{\tilde{\Sigma}} = E_{\Sigma} - c.
\end{equation}

The explicit form of the radial profile \(\tilde{r}\) follows from integration:
\begin{align}
    \frac{\tilde{r}'}{\tilde{r}} &= c s + d, \\
    \frac{\mathrm{d}\tilde{r}}{\tilde{r}} &= (c s + d) \mathrm{d}s, \\
    \ln \tilde{r} &= \frac{c}{2} s^2 + d s + f, \\
    \tilde{r} &= e^f e^{\frac{c}{2} s^2 + d s} \equiv \tilde{r}_0 e^{\frac{c}{2} s^2 + d s}.
\end{align}

For the original surface \(\Sigma\), the corresponding radial function is then given by:
\[
 r = \frac{1}{\tilde{r}} = r_0 e^{-\frac{c}{2} s^2 - d s}.
\]
\section{\textbf{Analytical solutions on the nanocone}}
We solve the rescaled eigenvalue equation from the main text
\begin{align}
  \label{eq_Dirac_super_SM}
    \begin{pmatrix}
			0 & \hat{L}^- \\
			\hat{L}^+ & 0
	\end{pmatrix}
    \psi_\Sigma(s)
    =
    E_\Sigma\psi_\Sigma(s) \, ,\quad \\
    \hat{L}^\pm = i\bigg( -\partial_s - \frac{1}{2}\frac{r'(s)}{r(s)}  \pm V_\Sigma(s)\bigg),
\end{align}
on the nanocone $r(s) = s \sin(\theta / 2)$.
First we rescale the wave function 
\begin{equation}
    \psi_\Sigma \to \frac{1}{\sqrt{r}} \psi_\Sigma, 
\end{equation}
which renders the differential operator 
\begin{equation}
\label{eq:rescaled_L}
\hat{L}^\pm \to \hat{L}^\pm = i\bigg( -\partial_s  \pm V_\Sigma(s)\bigg).
\end{equation}

Next, we explicitly calculate the wave function of the ground state.
To do so, we rewrite the effective mass potential as
	\begin{equation}
		V_\Sigma(s) = \frac{\alpha^\theta_m}{s} +  \beta_\theta s,
	\end{equation}
		where we defined the quantities
		\begin{equation}
		\beta_\theta \equiv {\sin(\theta / 2)}{\beta} \; \text{  and  } \; \alpha_m^\theta \equiv \frac{\alpha_m}{\sin(\theta / 2)}.
	\end{equation}
This definition will be useful as these variables will appear as exponents.
	Based on $V_\Sigma(s)$ we see that the ground state only exists if the magnetic field $B$ and the angular number $\alpha_m$ have different signs. By plugging in, it is easy to show that the ground state in the case $B > 0$ and $\alpha_m < 0$ is given by 
\begin{equation}
	\label{eq:zero_up}
	\psi_\Sigma^{0}(s) = \mathcal{N}_0 \, p_0(s) \, 
	\left( \begin{matrix}
		1 \\
		0
	\end{matrix} \right),
\end{equation}
	with the normalization constant $\mathcal{N}_0$ and 
	\begin{equation}
		\label{eq:cone_p_0}
		p_0(s) = s^{|\alpha_m^\theta |} \exp\Big\{-|\beta_\theta| s^2 / 2 \Big \},
	\end{equation}
	while in the other case $B < 0$ and $\alpha_m > 0$ we obtain
\begin{equation}
	\label{eq:zero_down}
	\psi_\Sigma^{0}(s) = \mathcal{N}_0 \, p_0(s) \,
	\left( \begin{matrix}
		0 \\
		1
	\end{matrix} \right).
\end{equation}
Having derived the expression for the ground state, we can proceed to analyze the explicit form of the excited states.
Using the rescaled form of the differential operator $\hat{L}^\pm$ from Eq. \eqref{eq:rescaled_L}, we are ready to write down the decoupled differential equation
 	 \begin{equation}
 	\label{eq:L_down_component}
	 	\hat{L}^-\hat{L}^+ \psi_\downarrow(s) = E_\Sigma^2 \psi_\downarrow
	 \end{equation} 
    for the excited states, where we order the terms by powers in $s$, namely
	\begin{equation}
		\label{eq:cone_equation}
		\bigg(-\partial^2_s + A_1 - E_\Sigma^2 + \frac{A_2}{s^2} + A_3 s^2\bigg) \psi_\downarrow(s) = 0,
	\end{equation}
	with the factors
	\begin{align}
		A_1 &= \beta_\theta + 2 \beta_\theta \alpha_m^\theta  \\
		A_2 &= \alpha_m^\theta(\alpha_m^\theta - 1)\\
		A_3 &= \beta_\theta^2.
	\end{align}
The most important properties to keep in mind are $A_3 > 0$ and $A_2 > - 1 / 4$.
	The following ansatz leads to an analytical solution
	\begin{equation}
		\label{eq:INFO}
		\psi_\downarrow(s) = \frac{\mathcal{N}}{\sqrt{s}} {p\Big(|\beta_\theta| s^2\Big)},
	\end{equation}
	with some new function $p(z)$ and normalization constant $\mathcal{N}$.
	After some calculation steps, one arrives at the two linearly independent solutions, which are the Whittaker functions
	\begin{equation}
    \label{eq:solution-linear-indep}
		p \Big(|\beta_\theta| s^2\Big) = M_{\kappa, \mu}\Big(|\beta_\theta|s^2 \Big) \quad \text{and} \quad W_{\kappa, \mu}\Big(|\beta_\theta|s^2\Big),
	\end{equation}
	with
	\begin{align}
		\kappa &= \frac{-A_1 + E_\Sigma^2}{4A_3} \, , \\
		\mu &= \frac{1}{2}\sqrt{A_2 + 1/4} = \frac{1}{2}|\alpha^\theta_m - 1/2| > 0. \notag
	\end{align}
	These Whittaker functions can be further expressed in terms of Kummer's confluent hypergeometric functions $M(a, b, z)$ and $U(a, b, z)$, with the variables $a, b, z \in \mathbb{R}$, by
\begin{align}
	\label{eq:whittaker_definition}
	M_{\kappa ;\mu }\left(z\right)&=\exp \left(-z/2\right)z^{\mu +{\tfrac {1}{2}}}M\left(\mu -\kappa +{\tfrac {1}{2}},1+2\mu ,z\right), \notag \\
	W_{\kappa ;\mu }\left(z\right)&=\exp \left(-z/2\right)z^{\mu +{\tfrac {1}{2}}}U\left(\mu -\kappa +{\tfrac {1}{2}},1+2\mu ,z\right). \notag \\
        & 
\end{align}
	Since we want the solutions to be bound states, we demand 
	\begin{align}
		\label{eq:limes_cone}
		\lim_{s\rightarrow 0} \psi_\downarrow(s) &\rightarrow 0, \\
		\lim_{s\rightarrow \infty} \psi_\downarrow(s) &\rightarrow 0. \notag
	\end{align}

	The first limit in Eq. \eqref{eq:limes_cone} is fulfilled by default since $\mu >  0$ in Eq. \eqref{eq:whittaker_definition} and by definition $M(a, b, 0) = 1$, leading to
	\begin{equation}
		\psi_\downarrow(0) = M_{\kappa; \mu}(0) = W_{\kappa; \mu}(0) = 0.
	\end{equation}
	Therefore, the only relevant case we have to worry about is $s \to \infty$. The only way to achieve a vanishing solution at infinity is by requiring the Kummer's functions $M(a, b, z)$ and $U(a, b, z)$ to be a polynomial, which is the case if the first argument is a non-positive number $-n$ where $n \in \mathbb{N}$, which will be the principal quantum number. In our case this amounts to
	\begin{equation}
		\label{eq:condition_energy_cone}
		\mu - \kappa + 1/2 \overset{!}{=} -n.
	\end{equation}
	The hypergeometric functions are then reduced to generalized Laguerre polynomials and become linearly dependent
	\begin{equation}
    \label{eq:func-condition}
		U(-n, 1 + 2\mu, z) \propto M(-n, 1 + 2\mu, z) \propto L_n^{ 2\mu}(z).
	\end{equation}
	By rearranging Eq. \eqref{eq:condition_energy_cone}, we arrive at the spectrum $E_\Sigma$ and notice that it splits into two categories, namely, for $\sgn(B) \neq \sgn(\alpha_m)$ we obtain
	\begin{equation}
		\label{eq:energies_cone_1_SM}
		E^g_\Sigma(\alpha_m, \beta, n) = \pm \sqrt{4 |\beta_\theta| n} \underset{B \to \infty}{\propto} B^{\frac{1}{2}},
	\end{equation}
	while for $\sgn(B) = \sgn(\alpha_m)$ we have
	\begin{equation}
		\label{eq:energies_cone_2_SM}
			E^f_\Sigma(\alpha_m, \beta, n) = \pm \sqrt{4|\beta_\theta|\Big (\frac{1}{2} + n + |\alpha^\theta_m| \Big )} \underset{B \to \infty}{\propto} B^{\frac{1}{2}}.
	\end{equation}
	In both cases, we asymptotically get a {square root dependence of the energies} as a function of magnetic field $B$.
	
	As we can see in the analytical expression, Eq. \eqref{eq:energies_cone_2_SM}, the conical geometry with $\theta < \pi$ leads to the breaking of the $m$-degeneracy of the standard Dirac-Landau levels from Eq. \eqref{eq:energies_cone_1_SM}. To the best of our knowledge, these novel energy levels for massless Dirac electrons on the cone have not been calculated before. A similar problem, yet only for Schr\"odinger electrons, has been addressed in \cite{Marques_2001}, where a degeneracy breaking can also be observed.

	Let us now take care of the wavefunctions and deliver the explicit form of the arc length spinor $\psi_\Sigma(s)$. Before starting, a short remark is necessary. We have already seen that the energies fall into two branches, namely for equal and opposite sign of $\alpha_m$ and $B$, and thus the absolute sign is not of importance, only the relative one. However, this is no longer the case for the wavefunctions. We have already seen in the calculation of the zero energy levels, Eqs. \eqref{eq:zero_down} and \eqref{eq:zero_up}, that the absolute sign of $B$ and $\alpha_m$ influences the form of the ground state wave function. The same behavior is to be expected in the calculation of the excited states. Therefore, we expect slightly different solutions for different absolute signs of $B$ and $\alpha_m$. Fortunately, we do not need to calculate all four cases, as we can make use of time reversal transformations $\mathcal{T}$ to translate solutions from $B$ and $\alpha_m$ to $-B$ and $\alpha_{-m}$, since time reversal transformations flip the sign of magnetic fields and angular momentum, and thus also the sign of the angular momentum quantum number. Therefore, we compute the wavefunctions only for two cases, namely for ($B < 0$, $\alpha_m>0$) and ($B>0$, $\alpha_m>0$) and then apply the time reversal operator, i.e.
	\begin{equation}
		\mathcal{T} = i \sigma_y \mathcal{K}
	\end{equation}
	on the spinor $\psi(s)$ to obtain the remaining two cases. Here we denote the complex conjugation operator by $\mathcal{K}$. 
    
	Let us start with $B < 0$ and $\alpha_m > 0$. 
    Using Eqs. \eqref{eq:solution-linear-indep}, \eqref{eq:whittaker_definition}, \eqref{eq:condition_energy_cone} and \eqref{eq:func-condition}, the ansatz, Eq. \eqref{eq:INFO},
    becomes
	\begin{equation}
	\psi_\downarrow(s) = \mathcal{N} p_0(s) L_n^{|\alpha_m^\theta| - 1 / 2}(|\beta_\theta| s^2),
\end{equation}
	using the function $p_0(s)$ from Eq. \eqref{eq:cone_p_0}.  The other component can easily be obtained by applying the differential operator $\hat{L}^+$ on the down-spin component, i.e. 
	\begin{equation}
		\label{eq:operator_up}
		\psi_\uparrow(s) = \frac{i(-\partial_s + V_\Sigma(s))}{E_\Sigma} \psi_\downarrow(s),
	\end{equation}
	which gives us the total spinor
	\begin{equation}
		\label{eq:excited_1}
		\psi_{\Sigma}(s) = \mathcal{N} p_0(s) \begin{pmatrix}
		{\frac{2i|\beta_\theta|}{|E^g_\Sigma(\alpha_m, \beta. n)|}} s L_{n-1}^{|\alpha_m^\theta| + 1 / 2}(|\beta_\theta| s^2) \\
				L_n^{|\alpha_m^\theta| - 1 / 2}(|\beta_\theta| s^2) 
		\end{pmatrix}.
	\end{equation}
	We observe that the up component has one polynomial order less than the down component, which is consistent with the structure of the ground state, see Eq. \eqref{eq:zero_down}.
	Performing a similar computation for the other case ($B >0$, $\alpha_m>0$) yields the total spinor
	\begin{equation}
		\label{eq:excited_2}
		\psi_{\Sigma}(s) = \mathcal{N} p_0(s) \begin{pmatrix}
			\frac{2i|\beta_\theta|}{|E^f_\Sigma(\alpha_m, \beta, n)|}sL_n^{|\alpha^\theta_m| + 1/2}(|\beta_\theta| s^2)\\ L_n^{|\alpha^\theta_m |- 1/2}(|\beta_\theta| s^2)
		\end{pmatrix},
	\end{equation}
	where this time we obtain an order higher in the up component. The effect of $\mathcal{T}$ preserves the overall shape of the states, only that the {up component is swapped with the down component}.

\section{\textbf{Local current density and magnetic moment}}
In this section, we calculate the probability current density $\Vec{j}$ on general curved and rotationally symmetric surfaces $\Sigma$, which are parameterized by arc length and azimuthal angle, $(s, \phi)$. We consider the physical radius of revolution, which is expressed as $\rho(s)$. 
Subsequently we calculate the magnetic moment on curved surfaces starting from $\Vec{j}$.

We start by specifying the Dirac equation in curved space
\begin{align}
\label{eq:initial_lagrangian}
    i \hbar \partial_t \psi_\Sigma = \hbar v_F \bigg[-i&\sigma_1 \bigg(\partial_s + \frac{\rho'(s)}{2\rho(s)}\bigg)+ \\ &\sigma_2 \bigg(-i\frac{\partial_\phi}{\rho(s)} + \frac{eB_z}{2\hbar} \rho(s)\bigg)\bigg] \psi_\Sigma, \notag
\end{align}
where $\sigma_i$ are the usual Pauli matrices.
A possible Lagrangian is given by
\begin{widetext}
    \begin{align}
    \mathcal{L}(t, s, \phi) = \sqrt{-\mathcal{G}} \hbar v_F \psi_\Sigma^\dagger  \bigg[-i\sigma_1 \bigg(\partial_s + \frac{\rho'(s)}{2\rho(s)}\bigg) + \sigma_2 \Big(-i\frac{\partial_\phi}{\rho(s)} + \frac{eB_z}{2\hbar} \rho(s)\Big) - i \frac{\partial_t}{v_F}\bigg] \psi_\Sigma.
\end{align}
\end{widetext}
The action is then calculated by
\begin{equation}
   S = \int \mathrm{d}t \int_\Sigma \mathrm{d}s \mathrm{d}\phi \, \mathcal{L}(t, s, \phi).
\end{equation}
In our case, for a general surface of revolution we obtain $\sqrt{-\mathcal{G}} = \rho(s)$.
The Lagrangian is invariant under the global \emph{U(1)} - transformation of the wavefunction
\begin{eqnarray}
\psi_\Sigma \to \exp(i \alpha) \psi_\Sigma  &=& \psi_\Sigma + i \alpha \psi_\Sigma + \mathcal{O}(\alpha^2) \\
&=:& \psi_\Sigma  + \alpha \Delta \psi_\Sigma  + \mathcal{O}(\alpha^2). \notag
\end{eqnarray}
This yields the conserved Noether current
\begin{equation}
 \mathcal{J}^{\mu }:={\frac {\partial{\mathcal {L}}}{\partial(\partial _{\mu }\psi_\Sigma ) }} \Delta \psi_\Sigma  +{\frac {\partial{\mathcal {L}}}{\partial(\partial _{\mu }\psi_\Sigma ^\dagger) }} \Delta \psi_\Sigma ^\dagger.
\end{equation}
We therefore obtain explicitly
\begin{align}
    \mathcal{J}^t &= \hbar \rho(s) \psi_\Sigma ^\dagger \psi_\Sigma  \\
    \mathcal{J}^\phi &= \hbar v_F \psi_\Sigma ^\dagger {\sigma_2} \psi_\Sigma  \\
    \mathcal{J}^s &= \hbar v_F \rho(s) \psi_\Sigma ^\dagger \sigma_1 \psi_\Sigma.
\end{align}
By writing down $\partial_\mu \mathcal{J}^\mu = 0$, one obtains
\begin{eqnarray}
0 &=& \partial_t \Big( \psi_\Sigma ^\dagger \psi_\Sigma \Big) \\
&& + v_F \frac{1}{\rho(s)} {\partial_\phi}\Big(\psi_\Sigma^\dagger {\sigma_2} \psi_\Sigma \Big) \notag \\
&& + v_F \frac{1}{\rho(s)} \partial_s \Big(\rho(s) \psi_\Sigma^\dagger \sigma_1 \psi_\Sigma\Big). \notag
\end{eqnarray}
The spatial divergence acting on a spatial tangential vector in the coordinates $(s, \phi)$ with $\mathcal{G}_{ss} = 1$, $\mathcal{G}_{\phi \phi} = \rho^2(s)$ is given by
\begin{equation}
    \nabla \Vec{j} = \frac{1}{\rho(s)} \partial_\phi j_\phi + \frac{1}{\rho(s)} \partial_s (\rho(s) j_s).
\end{equation}
Identifying the charge density $\varrho= -e \psi^\dagger \psi$, we easily obtain the continuity equation
\begin{equation}
    \frac{\partial \varrho}{\partial t} + \nabla \vec{j} = 0,
\end{equation}
with (spatial) current components
\begin{align}
    j_\phi &= -ev_F \psi_\Sigma^\dagger \sigma_2 \psi_\Sigma, \\
    j_s &= -ev_F \psi_\Sigma^\dagger \sigma_1 \psi_\Sigma.
\end{align}
We proceed and calculate the magnetic moment in $z$ direction which reads
\begin{equation}
    m_z = \frac{1}{2}\int_\Sigma \mathrm{d}s \mathrm{d}\phi \rho(s) \bigg(\Vec{\rho} \times \Vec{j}\bigg)_z,
\end{equation}
where 
\begin{equation}
    \vec{\rho} = \begin{pmatrix}
\rho(s) \cos(\phi) \\ \rho(s) \sin(\phi)\\
z(s)
\end{pmatrix}.
\end{equation}
We express the current density in the local coordinates $\vec{j} = j_\phi \mathbf{e}_\phi + j_s \mathbf{e}_s$, with the tangential directions $\mathbf{e}_\phi$ and $\mathbf{e}_s$ given by 
\begin{equation}
   \mathbf{e}_\mu = \frac{\partial_\mu \vec{\rho}}{|\partial_\mu \vec{\rho}|}.
\end{equation}
We realize that the surviving part in the cross product is
\begin{align}
\label{eq:magnetization}
    m_z = \frac{1}{2}\int_\Sigma \mathrm{d}s \mathrm{d}\phi \rho^2(s) j_\phi = -\frac{e v_F}{2}\int_\Sigma \mathrm{d}s \mathrm{d}\phi \rho^2(s) \psi_\Sigma^\dagger \sigma_2 \psi_\Sigma.
\end{align}

Note that by utilizing the Hellman-Feynman theorem, it is easy to show that  $m_z$ is consistent with the thermodynamical definition
\begin{equation}
    m_z = -\frac{\partial E_\Sigma}{\partial B}.
\end{equation}

\section{\textbf{Vanishing magnetic moment for the geometry-sensitive branch}}
We are now prepared to analytically compute the magnetic moment \( m_z \) on the reciprocal cone for the states with \( \text{sgn}(\alpha_m) \neq \text{sgn}(B) \), corresponding to the geometry-sensitive branch. Using Eq. \eqref{eq:magnetization}, we have for the magnetization of the reciprocal surface $\tilde{r}(s) = 1 / r(s)$ the proportionality 
\begin{equation}
    m_z \propto \int_{s_0}^{s_1} \mathrm{d}s \tilde{r}^2(s) \psi_{\tilde{\Sigma}}^{\{\alpha_m;\beta\}\dagger} \sigma_2 \psi_{\tilde{\Sigma}}^{\{\alpha_m;\beta\}},
\end{equation}
where for simplicity we chose
$s_0 = 0$ and $s_1 = \infty$. As in the main paper, the corresponding quantum number $\alpha_m$ and dimensionless magnetic flux $\beta$ are specified as superscripts.
Using the definition of the cone profile $r(s) = s \sin(\theta / 2)$ and the theorem from the main text, $\psi_{\tilde{\Sigma}}^{\{\alpha_m;\beta\}} = r(s) \psi_{{\Sigma}}^{\{\beta;\alpha_m\}}$, we obtain the magnetic moment for the geometry-sensitive states on the reciprocal cone

\begin{align}
\notag
     m_z &\propto \int_0^\infty \mathrm{d}s \frac{1}{s}p_0^2(s) s L_{n-1}^{{|Y|}+ 1 / 2}(|X| s^2)L_n^{|Y| - 1 / 2}(|X| s^2) \\ &= \int_0^\infty \mathrm{d}s p_0^2(s) L_{n-1}^{{|Y|}+ 1 / 2}(|X| s^2)L_n^{|Y| - 1 / 2}(|X| s^2),
\end{align}

where 
\begin{equation}
    X \equiv  \sin(\theta / 2) \alpha_m \quad \text{and} \quad Y \equiv \frac{\beta}{\sin(\theta / 2)}.
\end{equation}
To solve the integral we choose the transformation $z =|X| s^2$,
which gives
%\begin{widetext}
\begin{align}
     m_z &\propto \int_0^\infty \frac{\mathrm{d}z}{\sqrt{z}} z^{|Y|} \exp(-z) L_{n-1}^{{|Y|}+ 1 / 2}(z)L_n^{|Y| - 1 / 2}(z)
     \\ &= \int_0^\infty {\mathrm{d}z} z^{|Y|- 1/2} \exp(-z) L_{n-1}^{{|Y|}+ 1 / 2}(z)L_n^{|Y| - 1 / 2}(z). \notag
\end{align}
%\end{widetext}
We further use an identity for generalized Laguerre polynomials from literature, namely
\begin{equation}
    L_{n}^{(\alpha + 1)}(x) = \sum_{i=0}^n L_i^{(\alpha)}(x),
\end{equation}
which gives us
\begin{equation}
 m_z \propto \sum_{i=0}^{n-1}\int_0^\infty {\mathrm{d}z} z^{|Y|- 1/2} \exp(-z) L_{i}^{{|Y|} - 1 / 2}(z)L_n^{|Y| - 1 / 2}(z).
\end{equation}
We then use the orthogonality relation, resulting in 
\begin{align}
 m_z &\propto \sum_{i=0}^{n-1}\int_0^\infty {\mathrm{d}z} z^{|Y|- 1/2} \exp(-z) L_{i}^{{|Y|} - 1 / 2}(z)L_n^{|Y| - 1 / 2}(z) \notag
 \\
 &= \sum_{i=0}^{n-1} \frac{\Gamma(n + |Y| - 1 / 2 + 1)}{n!} \delta_{ni} = 0.
\end{align}
We have therefore analytically demonstrated that the magnetic moment on the reciprocal cone vanishes for the geometry-sensitive states, where \( \text{sgn}(\alpha_m) \neq \text{sgn}(B) \).

\section{\textbf{Asymptotic behavior of transcendental surfaces}}

In this paragraph we are interested in the numerical behavior of the spectrum of the transcendental surface

\begin{equation}
    \frac{\mathrm{d}r}{\mathrm{d}s} \propto \sin(r),
\end{equation}

whose unique radius solution reads
\begin{equation}
    r(s) = 2 \arctan(\exp(s)).
\end{equation}

\begin{figure}[H]
    \centering
\includegraphics[width=\linewidth]{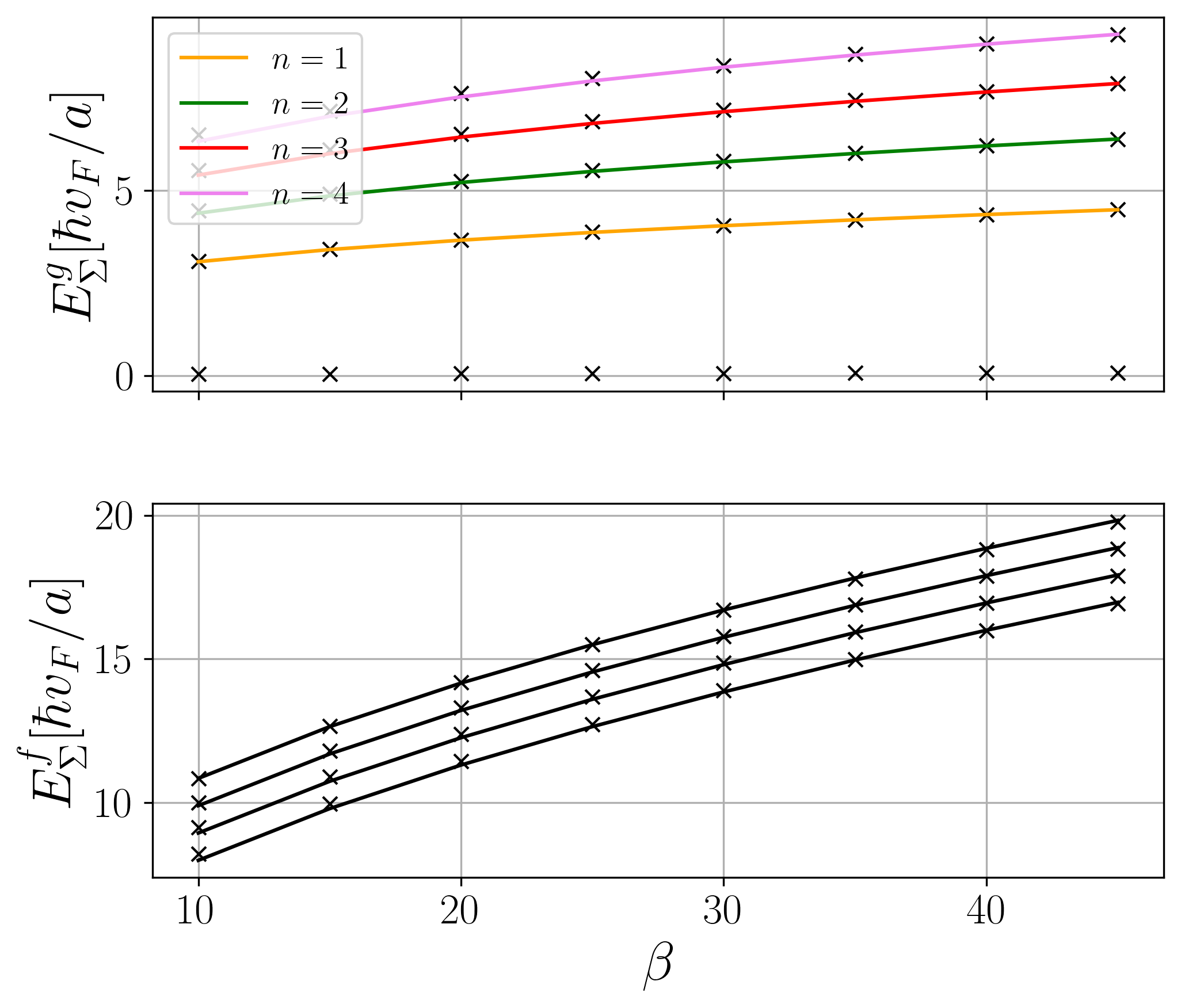}
        \footnotesize
       \caption{Energy as a function of flux of the radial profile with $r' \propto \sin(r)$ for a fixed angular quantum number $\alpha_m$ in each case ($m=-1$ for the upper figure and $m=1$ for the lower figure). The radial profile equation is solved, and its spectrum is computed numerically, which is shown by the crosses. In the computations we used an arc length lattice spacing of $\Delta_s = 0.225[a]$. The continuous lines represent the fitting polynomial $E(\beta) = q\beta^d + \delta$ with the variables having the values: (upper) $q = \textcolor{orange}{\sqrt{3.0}}, \textcolor{green}{\sqrt{6.1}}, \textcolor{red}{\sqrt{9.3}}, \textcolor{purple}{\sqrt{12.7}}$, $\delta=0$, $d = 1 / 4$ 
       and (lower) $q = \sqrt{6.4}$, $d = 1 / 2$, $\delta = 0.95,\, 1.9,\, 2.85$ and $\delta = 3.8$, confirming our conjecture.}
    \label{sinus-fig}
\end{figure}

Fig.~\ref{sinus-fig} shows the spectrum of this surface for a given angular quantum number $m$ and different main quantum numbers $n$ under a varying magnetic field. As can be seen from the fitting curves, the same sign branch $\sgn(B) = \sgn(\alpha_m)$ scales with the square root, while the different sign branch $\sgn(B) \neq \sgn(\alpha_m)$ scales with the fourth root. This confirms our conjecture, since 
\begin{equation}
    \sin(r) = r^1 - \frac{r^3}{3!} + \dots,
\end{equation}
and hence $\eta_\Sigma =(2 - N) / 4 = 1/4$ should be the dominant contribution.

\end{document}